# Exceso de muertes oculto a 100 días de la cuarentena en Perú por COVID-19

Ronal Arela-Bobadilla[a, 1]

[a] *Universidad Católica San Pablo, Arequipa, Perú.*

**Resumen**

**Objetivo:** Realizar una estimación del exceso de muertes ocasionado por COVID-19 en la mortalidad no violenta del Perú controlando el efecto de la cuarentena.

**Métodos:** Análisis de datos longitudinales de los departamentos del Perú utilizando información pública oficial del Sistema Informático Nacional de Defunciones y del Ministerio de Salud del Perú. El análisis se realiza entre el 1 de enero de 2018 y el 23 de junio de 2020 (100 días de cuarentena). Se ha utilizado la tasa de muertes diarias por millón de habitantes. Para la estimación del impacto de la cuarentena se utilizaron los días en los que los departamentos permanecieron en cuarentena con una cantidad límite de casos de COVID-19 acumulados. Se establecieron tres límites para los casos: menos de 1, 10 y 100 casos.

**Resultado:** En Perú, la tasa diaria de muertes por millón de habitantes se redujo en -1,89 (95%CI: -2,70; -1,07) en los días de cuarentena y sin casos de COVID-19. Al comparar este resultado con el número de total de muertes no violentas, el exceso de muertes durante los 100 primeros días de cuarentena es de 36 230. Esta estimación es 1,12 veces la estimada con datos de 2019 y 4,2 veces las muertes oficiales por COVID-19.

**Conclusión:** La cuarentena redujo las muertes no violentas; sin embargo, son opacadas por el incremento a causa directa o indirecta de la pandemia. Por lo tanto, la diferencia entre el número de fallecimientos actuales y el de años pasados subestima el exceso real de defunciones.

---

[1] Email: rwarela@ucsp.edu.pe

**Palabras clave:** COVID-19, Mortalidad, Registros de Mortalidad, Cuarentena.

El primer caso de COVID-19 en Perú se registró el 6 de marzo de 2020 en Lima. A partir de ese momento, el Gobierno tomó una serie de acciones para contener la propagación. El 11 de marzo mediante el Decreto Supremo N°008-2020-SA se declaró el estado de emergencia sanitaria a nivel nacional por 90 días y la postergación de las labores escolares en los colegios públicos hasta el 30 de marzo (1). El día 12 del mismo mes se suspenden las actividades presenciales en centros de educación superior públicos y privados (2). Un hito importante sucede el 15 de marzo, día en el que se aprueba el Decreto Supremo N°044-2020-PCM que declara el estado de emergencia nacional por 15 días y establece el aislamiento social obligatorio o cuarentena en todo el país; sin embargo, se exceptuaron algunas actividades esenciales como la producción de alimentos (3). En el decreto también se suspende el transporte interprovincial de pasajeros por medio terrestre, aéreo y fluvial. A continuación, se produjo una serie de medidas que alargaron la cuarentena hasta el 30 de junio de 2020 (4); aunque con algunas variantes que las hicieron más estrictas.

El cumplimiento de la cuarentena fue resguarda por las fuerzas militares y policiales y se ha mantenido a pesar de algunos casos de incumplimiento. Los datos de Google sobre movilidad local muestran un acatamiento sostenido en el país. La reducción de la movilidad de personas se ha reducido entre -77% en tiendas y lugares de ocio y -44% en supermercados y farmacias hasta el 29 de mayo de 2020 (5).

Esta situación ha provocado la muerte de 8 616 personas hasta el 23 de junio de 2020 –día 100 de la cuarentena nacional– (6); sin embargo, el Sistema Informático

Nacional de Defunciones (SINADEF) registra un número muy superior de fallecimientos por causa no violenta en comparación al mismo periodo de 2019 (7). Entre el 16 de marzo y el 23 de junio del presente año se han registrado 32 251 decesos más que entre las mismas fechas de 2019. Aun quitando los 8 616 registros oficiales a causa del COVID-19, hay una anomalía de 23 635 defunciones.

Sin embargo, las comparaciones del registro de muertes actual con el de años anteriores puede generar estimaciones incorrectas, debido a la situación de cuarentena de la población en 2020 (8) o a otros factores diferentes entre ambos años (9). Por lo tanto, la diferencia entre la mortalidad observada en 2020 y la de años anteriores es un estimador sesgado del impacto perjudicial que ha tenido la situación originada por el COVID-19.

La presente investigación realiza una estimación del exceso de muertes ocasionado por COVID-19 en la mortalidad no violenta del Perú controlando el efecto de la cuarentena. Esto propone que la mortalidad ha sufrido dos cambios que no son visibles individualmente. El primero es una reducción de las muertes no violentas debido al confinamiento obligatorio y el segundo es un incremento de este tipo de decesos debido a la crisis. Por lo tanto, el exceso de muertes registradas entre 2020 en comparación a los años pasados subestima el verdadero impacto negativo de la crisis.

**Materiales y métodos**

**Diseño del estudio y fuentes de datos.**

La investigación utilizó información pública disponible en las entidades correspondientes del país. De esta forma, no ha sido necesaria la participación de un comité de ética.

Se realizó un análisis longitudinal de los 24 departamentos del Perú, cuyas unidades de tiempo fueron los días entre el 01 de enero de 2018 y el 23 de junio de 2020.

Para cada día en cada departamento el número de observaciones no ha sido el mismo debido a que algunos departamentos tuvieron su primer caso de COVID-19 en días diferentes –datos panel desbalanceado (Tabla 1)–.

| Tabla 1. Muertes diarias por millón de habitantes promedio según periodo, Perú, 2020 | | | | |
|---|---|---|---|---|
| Departamento | Antes de la cuarentena[a] (ppmd) | Después del inicio de la cuarentena y antes del caso 1 de Covid-19 (ppmd) | Después del caso 1 de Covid-19 hasta el día 100 de la cuarentena[b] (ppmd) | Casos Covid-19 al inicio de la cuarentena (casos) |
| Amazonas | 7,0 | 4,2 | 6,1 | 0 |
| Ancash | 13,5 | NA | 23,3 | 1 |
| Apurímac | 10,0 | 7,7 | 11,7 | 0 |
| Arequipa | 11,6 | NA | 16,0 | 2 |
| Ayacucho | 8,7 | 5,8 | 5,7 | 0 |
| Cajamarca | 6,7 | 4,2 | 5,7 | 0 |
| Cusco | 13,4 | NA | 10,8 | 1 |
| Huancavelica | 14,2 | 11,3 | 15,5 | 0 |
| Huánuco | 10,5 | NA | 11,4 | 2 |
| Ica | 13,1 | NA | 26,3 | 1 |
| Junín | 12,0 | 11,8 | 13,9 | 0 |
| La Libertad | 11,5 | NA | 18,8 | 1 |
| Lambayeque | 5,6 | NA | 11,7 | 3 |
| Lima[c] | 9,7 | NA | 28,8 | 73 |
| Loreto | 6,3 | NA | 20,0 | 0 |
| Madre de Dios | 13,6 | 14,2 | 21,2 | 0 |
| Moquegua | 13,3 | 12,4 | 12,3 | 0 |
| Pasco | 8,4 | 13,3 | 11,0 | 0 |
| Piura | 8,6 | NA | 19,8 | 2 |
| Puno | 12,1 | 8,2 | 8,6 | 0 |
| San Martin | 9,2 | 7,4 | 10,6 | 0 |
| Tacna | 10,6 | 5,8 | 7,2 | 0 |
| Tumbes | 12,2 | 13,3 | 32,8 | 0 |
| Ucayali | 10,6 | 10,2 | 30,7 | 0 |
| **Total** | **10,6** | **8,8** | **16,2** | **86** |

Fuente: Análisis del autor en base a los datos del Sistema Informático Nacional de Defunciones (SINADEF) y del Ministerio de Salud (MINSA) y del XII Censo Nacional de Población y Vivienda 2017 del Instituto Nacional de Estadística e Informática (INEI).

[a] La cuarentena se inicia el día 16 de marzo de 2020.

[b] El día 100 de la cuarentena se cumple el 23 de junio de 2020.

[c] Incluye la provincia constitucional del Callao.

NA: No se aplica

ppmd: personas por millón al día.

Cada unidad de análisis se define como una zona geográfica separada del resto. En este sentido, la provincia constitucional del Callao, que cuenta con un gobierno regional propio, y el departamento de Lima se agruparon en una sola categoría debido a su cercanía. Su agrupación no ha tenido impactos trascendentales en las estimaciones.

La información de las muertes no violentas ocurridas entre el 01 de enero de 2018 y el 3 de junio de 2020 se obtuvo del SINADEF. En el sistema se registran las defunciones generales, fetales y de personas no identificadas (10).

El SINADEF se implementó en 2016 en Perú como respuesta a la baja cobertura de las defunciones y sus causas en años anteriores (11). Desde su implementación, la calidad del registro se ha incrementado; sin embargo, aún no presenta una cobertura total (11,12). El SINADEF registró el 75,49% de las defunciones registradas por el Instituto Nacional de Estadística e Informática (INEI) en 2018 que se basó en los datos del Registro Nacional de Identificación y Estado Civil (RENIEC) (13). Solo se consideró la información desde 2018 debido a que es el último registro sobre su desempeño y tiene una clasificación de muertes violentas y no violentas mejor que en 2017.

La consecuencia directa de la falta de cobertura es una subestimación de las defunciones: se registran menos muertes que las reales. Esta situación no generó sesgos importantes en las estimaciones realizadas bajo el supuesto de que la varianza de las muertes por millón de habitantes no ha cambiado –homocedasticidad– (14).

La población censada de cada departamento se obtuvo del XII Censo Nacional de Población y Vivienda 2017. El registro se realizó en octubre de 2017 y es la información disponible más confiable y actualizada sobre el tamaño de la población a nivel nacional. Para asegurar la calidad de la información, se decretó una inmovilización parcial en el área urbana del país con el fin de eliminar la duplicación y omisión (15).

Los cambios poblacionales anuales desde 2007 no han sido grandes. Entre 2007 y 2017 la población nacional creció en promedio 0,7% al año. Al interior del país, 22 de los 24 departamentos registraron un incremento menor al 2% anual. Solo Madre de Dios registró un incremento de 2,6% y Huancavelica una reducción de -2,7% (16). En consecuencia, se espera que los cambios en la población entre 2017 y 2020 no hayan producido problemas en las estimaciones.

Los registros de los casos positivos por COVID-19 en cada departamento del Perú se recogieron de la Plataforma Nacional de Datos Abiertos del Estado Peruano disponible hasta el 23 de junio de 2020. La información es generada por el Ministerio de Salud (MINSA).

La información de los días de inicio de la cuarentena fue obtenida de los decretos supremos emitidos por el Gobierno y que se encuentran disponibles en el Diario Oficial El Peruano (3).

**Variables y estrategia de identificación**

La variable dependiente es el número de muertes diarias por causas no violentas por millón de habitantes en cada uno de los 24 departamentos del Perú entre el 01 de enero de 2018 y el 23 de junio de 2020.

Las muertes no violentas en Perú correspondieron al 94,65% de las muertes totales a nivel nacional entre el 01 de enero de 2018 y el 23 de junio de 2020.

La variable independiente que indica la situación de cuarentena es binaria e igual a 1 en los días de cuarentena –a partir del 16 de marzo de 2020 (3)– y 0 en los días sin cuarentena –antes del 16 de marzo de 2020–.

La estrategia para la identificación del impacto de la cuarentena ha sido la restricción de la muestra a los días antes de la cuarentena (cuarentena=0) y los días después de esta (cuarentena=1) en los que los departamentos tuvieron menos de una cantidad límite de casos positivos de COVID-19.

Se definieron 3 tipos de muestra en base a 3 cantidades límite de casos: la muestra tipo L1 considera a los días de los departamentos antes de su primer caso –casos acumulados menores a 1– la muestra tipo L10 antes de los 10 primeros casos y la muestra tipo L100 antes de los 100 primeros casos. La muestra L100 es la más grande y contiene a L10 y ésta a su vez contiene a L1.

La definición de 3 límites nos permite tomar en cuenta diferentes retrasos en el impacto negativo de las muertes originadas de forma directa o indirecta por el COVID-19. Un límite muy restrictivo como el de la muestra L1 puede mejorar el insesgamiento de la estimación; sin embargo, tiene un tamaño de muestra pequeño. Por otro lado, a medida que la cantidad límite se incrementa (muestra L100), la muestra de días con cuarentena es mayor; sin embargo, la probabilidad de sesgo producido por el incremento de las muertes durante la pandemia aumenta.

**Análisis estadístico**

En primer lugar, se verificó la estacionariedad de cada una de las series de tiempo de las muertes por millón de habitantes en cada departamento. Se utilizó la prueba propuesta por Choi (17) para datos panel bajo la especificación de Dickey-Fuller. Los resultados indican que la tasa de mortalidad se ha comportado de forma estacionaria; es decir, sin tendencia ni patrones estacionales ($P$=0,0000).

Por otro lado, se realizó la prueba de autocorrelación serial según la propuesta de Wooldridge (18). El resultado indica fuerte evidencia en contra de correlación serial (*P*=0,7426).

El modelo que se estimó fue:

$$y_{it} = B_0 + B_1 x_{it} + u_i + e_{it} \ \forall \ Nc19_{it} < L; \ L = \{1; 10; 100\}$$

Donde $y_{it}$ son las muertes no violentas en el departamento $i$ en el día $t$; $x_{it}$ es una variable binaria igual a 1 si el departamento $i$ estuvo en cuarentena en el día $t$; $u_i$ son los factores idiosincráticos de cada departamento, $e_{it}$ es el término de error y $Nc19_{it}$ es el número total de casos de COVID-19 en el día $t$ y departamento $i$ y $L$ es el límite que toma los valores de 1 para la muestra L1, 10 para la muestra L10 o 100 para la muestra L100.

Luego, se compararon las estimaciones por efectos fijos y aleatorios con la prueba de endogeneidad de Hausman (19). No se encontraron diferencias estadísticamente significativas entre las estimaciones por efectos fijos y aleatorios (Muestra tipo L1: *P*=0,9819; Tipo L10: *P*=0,6924; Tipo L100: *P*=0,7004) favoreciendo a la estimación más eficiente por efectos aleatorios.

En la Tabla 2 se resumen las estimaciones para cada una de las tres muestras.

**Resultados**

En la muestra L1 se utilizaron 14 paneles, en la muestra L10 se utilizaron 23 y en la L100 se utilizaron los 24 paneles presentes en el país. Cada departamento tiene un número diferente de días debido a 2 motivos: primero, cada departamento sobrepasó cada uno de los tres límites establecidos en fechas diferentes y, segundo, existen departamentos que no registraron muertes en la totalidad de días del periodo analizado. En promedio, cada departamento tuvo 802 días en los que reportó al menos una defunción entre el 01

de enero de 2018 y el día que reportó el caso 100 de COVID-19. El departamento con más días tuvo 849 y el de menos días tuvo 637.

Las estimaciones se realizaron con un total de 10 691 observaciones para la muestra L1, 18 039 para la muestra L10 y 19 224 para la muestra L100 (Tabla 2).

La media de muertes por millón de habitantes en los departamentos fue de 10,6 (rango intercuartil [IQR]: 6,4). La Tabla 1 resume las muertes promedio por millón de habitantes para cada departamento.

Los resultados muestran una reducción de la tasa de mortalidad no violenta por millón de habitantes en cada uno de los departamentos en los días en los que hubo cuarentena, pero no se registraron casos de COVID-19 (Tabla 2).

Estas estimaciones muestran que la cuarentena, sin la intervención del COVID-19, logró reducir la tasa de mortalidad en hasta -1,89 personas por millón de habitantes (95%CI: -2,70; -1,07). Para la muestra con límites de 10 y 100 casos, la estimación del impacto se reduce a -1,78 (95%CI: -2,23; -1,32) y -1,45 (95%CI: -1,76; -1,13) respectivamente; sin embargo, mantiene su significancia estadística.

| Tabla 2. Estimación de los modelos según muestra, Perú, 2020 | | | |
|---|---|---|---|
| Muestra | L1 | L10 | L100 |
| Cuarentena | -1,89* | -1,78* | -1,45* |
|  | (0,000) | (0,000) | (0,000) |
| Constante | 10,62* | 10,55* | 10,52* |
|  | (0,000) | (0,000) | (0,000) |
| Observaciones | 10 691 | 18 039 | 19 224 |
| Departamentos | 14 | 23 | 24 |
| Rho | 0,19 | 0,25 | 0,25 |

Fuente: Análisis del autor a partir de los datos recogidos.

Nota: Estimación con datos de panel de efectos aleatorios.

* Significativo al 0.001 de significancia.

Pvalores entre paréntesis

Rho: Fracción de la varianza total de $u_i + e_{it}$ debido a $u_i$

Considerando las poblaciones de cada departamento, según estas estimaciones durante los primeros 100 días de la cuarentena (16 de marzo-23 de junio de 2020) y en ausencia de COVID-19 el número de muertes en el país debería de ubicarse entre 23 809 (estimador L1) y 25 077 (estimador L100). Durante el mismo periodo (a partir del 16 de marzo y el 23 de junio) en 2019 el número de muertes registradas por el SINADEF fue de 27 788. En 2020, el registro de muertes totales en el país en el mismo periodo fue de 60,039. Con estas estimaciones, el exceso de muertes total en el país llegaría a 36 230. La Tabla 3 muestra las estimaciones para cada departamento.

| Tabla 3. Estimación del excedente de muertes, Perú, 2020 ||||
| Departamento | Muertes en los primeros 100 días de cuarentena[a] (ppmd) | Exceso de defunciones según estimador L1[b] (personas) | Exceso de muertes según mismo periodo de 2019 (personas) | Muertes oficiales por Covid-19 en los 100 días de cuarentena (personas) |
| --- | --- | --- | --- | --- |
| Amazonas | 190 | 27 | - 36 | 34 |
| Ancash | 2 540 | 1 282 | 1 163 | 437 |
| Apurímac | 447 | 122 | 64 | 14 |
| Arequipa | 2 238 | 900 | 488 | 203 |
| Ayacucho | 339 | - 79 | - 220 | 21 |
| Cajamarca | 752 | 101 | - 136 | 54 |
| Cusco | 1 300 | - 90 | - 221 | 13 |
| Huancavelica | 518 | 92 | 56 | 10 |
| Huánuco | 818 | 200 | 71 | 47 |
| Ica | 2 285 | 1 330 | 1 151 | 433 |
| Junín | 1 721 | 456 | 270 | 116 |
| La Libertad | 3 354 | 1 650 | 1 352 | 496 |
| Lambayeque | 1 329 | 987 | 1 179 | 718 |
| Lima | 31 903 | 23 725 | 22 616 | 4 513 |
| Loreto | 1 770 | 1 381 | 1 212 | 332 |
| Madre de Dios | 244 | 116 | 88 | 24 |
| Moquegua | 181 | 3 | - 40 | 9 |
| Pasco | 257 | 133 | 122 | 20 |
| Piura | 3 732 | 2 486 | 2 151 | 715 |
| Puno | 1 005 | - 197 | - 405 | 16 |
| San Martin | 853 | 265 | 176 | 126 |
| Tacna | 203 | - 76 | - 149 | 8 |
| Tumbes | 707 | 496 | 455 | 112 |
| Ucayali | 1 353 | 921 | 844 | 145 |
| **Total** | **60 039** | **36 230** | **32 251** | **8 616** |

Fuente: Preparado por el autor en base a los resultados obtenidos en base a los datos del Sistema Informático Nacional de Defunciones (SINADEF) y del Ministerio de Salud (MINSA) y del XII Censo Nacional de Población y Vivienda 2017 del Instituto Nacional de Estadística e Informática (INEI).

[a] Los primeros 100 días de cuarentena corresponde al periodo entre el 16 de marzo y el 23 de junio de 2020.

[b] Algunos departamentos no reportaron muertes todos los días. Se utilizó la misma cantidad de días que reportaron defunciones en 2019.

ppmd: personas por millón al día.

Algunos departamentos tienen un exceso negativo (Ayacucho, Cusco, Puno y Tacna) debido a que la cantidad de muertos durante la cuarentena fue aún menor que la calculada utilizando los estimadores L1, L10 y L100.

La comparación simple con el año 2019 indica que el exceso de fallecimientos durante el 16 de marzo y el 23 de junio de 2020 fue de 3,7 veces las muertes oficiales por COVID-19 hasta esa fecha; sin embargo, las estimaciones considerando la cuarentena indican que el excedente fue entre 4,1 (muestra L100) y 4,2 veces (muestra L1).

**Discusión.**

Los resultados muestran que la cuarentena ha logrado disminuir el número de fallecimientos en el Perú en hasta -1,89 muertes por millón de habitantes al día que, sin embargo, son opacadas por el incremento producido por la crisis. La estimación del exceso de muertes totales durante la pandemia sería de hasta 36 230 personas (Tabla 3), esto es 4,2 veces las muertes oficiales producidas por COVID-19 en el país hasta el 23 de junio de 2020 y 1,12 veces el exceso calculado con los datos de 2019. Por lo tanto, la comparación del número de muertes actuales con la de años anteriores subestima el impacto negativo directo e indirecto de la pandemia.

Este trabajo propone una forma diferente de estimación del excedente de las muertes producidas durante la pandemia controlando la reducción de las defunciones debido al confinamiento de los hogares. La respuesta rápida y generalizada del gobierno peruano y la ausencia, bajo datos oficiales, de casos de COVID-19 en varios lugares del país se han aprovechado para la identificación de una estrategia de estimación. Otras investigaciones utilizan los periodos anteriores como referencia para el cálculo de un contrafactual adecuado (20) en un contexto diferente al de Perú y otros han controlado los impactos de las muertes por neumonía e influenza y descuentan estos a la totalidad de muertes observadas durante la pandemia (9). Sin embargo, el trabajo de Nogueira, Dee Araújo Nobre, Nicola, Furtado y Vaz Carneiro (8) es uno de los pocos que toma en cuenta la reducción de muertes debido al confinamiento.

Varios trabajos en los últimos meses han mostrado los impactos de la cuarentena en el control de la pandemia por COVID-19 específicamente; sin embargo, sus resultados son diferentes para cada contexto (21). Mientras algunas investigaciones encuentran evidencia a favor de una reducción en el número de muertes por COVID-19 (22–25) en otros no se ha logrado confirmar esta relación (26). En vista de esto, se plantea que una explicación posible para la reducción de muertes no relacionadas directamente a COVID-19 es que, de haber existido un impacto positivo de la cuarentena, debería esperarse que el confinamiento haya logrado reducir también el riesgo de muerte por otro tipo de enfermedades. Otras posibles causas a las que puede adjudicarse la reducción de muertes son los efectos positivos en la calidad del aire debido al confinamiento (27).

Asimismo, es posible que el confinamiento pueda tener un impacto en forma de 'u', con una reducción inicial de los decesos y un incremento en los días siguientes. Este comportamiento podría ser consecuencia del surgimiento de patologías asociadas al aislamiento social (28, 29). Sin embargo, a los 100 días de cuarentena, algunos departamentos han continuado con una tasa de mortalidad reducida (Tabla 1).

Por otro lado, entre las posibles causas de este exceso de muertes podría encontrarse la saturación de los servicios de salud, miedo de las personas por atenderse en las clínicas debido a posibles contagios, falta de pruebas de COVID-19 y la agudización de otro tipo de enfermedades y la aparición patologías ocurridas durante la cuarentena.

Este trabajo permite la comparación del exceso de muertes producido en Perú con el de otros países que no han impuesto una cuarentena o cuyo confinamiento empezó después de la aparición de los primeros casos locales de COVID-19. Asimismo, puede ayudar a identificar lugares en cuarentena que tengan ya un exceso de muertes no observado directamente debido a que el número de defunciones es similar al de años anteriores.

**Fortalezas.**

Una de las fortalezas de este trabajo es la estrategia de identificación para la estimación del impacto de la cuarentena sin el impacto del COVID-19. La disposición de la cuarentena durante los primeros días fue igual para todos los departamentos del Perú. Esta imposición del gobierno puede entenderse como un hecho exógeno a cualquier característica a priori de los departamentos, permitiendo la estimación correcta del impacto de la cuarentena. Solo algunas regiones del mundo han tenido estas características y pocas investigaciones, a búsqueda del autor, han tenido este enfoque.

Otro punto que ha permitido la estimación ha sido que los decesos en el Perú no presentan un patrón estacional. Según los datos reportados por el SINADEF, entre el periodo entre 2018 y el inicio de la cuarentena en 2020, el número de fallecimientos ha seguido un patrón estacionario sin mayores cambios. En todo el país la tasa de muertes por millón de habitantes se ha mantenido cercana a 10,6 antes del inicio de la cuarentena (Tabla 1).

Por otro lado, luego de la cuarentena la reducción de las muertes parece mantenerse en el tiempo. En lugares como Puno, la tasa de fallecimientos no ha sufrido grandes cambios desde el descenso registrado al inicio de la cuarentena.

**Limitaciones.**

La principal limitación de este trabajo es que los registros del SINADEF pueden presentar deficiencias en el conteo del número de muertes a nivel nacional. En 2018 se registró el 75,49% de las defunciones calculadas por otras instituciones del estado (13). Sin embargo, la solidez de las estimaciones se basó en que el comportamiento de las muertes es homocedástico en el tiempo y las pruebas de estacionariedad utilizadas sugieren ello. Si este supuesto se mantiene, las estimaciones presentadas, no sufrirían de sesgo (14). No obstante, si debido a la pandemia el registro ha sufrido deficiencias en el conteo, la brecha podría ser mayor, en vista de que se identificaron menos decesos que los reales. No se tuvo información sobre esta situación.

En todo el territorio peruano no se había observado un exceso de muertes entre enero de 2018 y marzo de 2020; no obstante, no debería dejarse de lado la posibilidad de que un incremento debido a la agudización de los brotes de Dengue haya coincidido con el brote de COVID-19 (30). Esta enfermedad ya afectaba al departamento de Cusco antes del inicio de la pandemia. Es posible que sus impactos se hayan sumado a los producidos por la pandemia.

**Conclusión.**

El confinamiento ha logrado reducir las muertes por millón de habitantes; sin embargo, esta reducción no es observable directamente debido a que el incremento de las muertes producidas de forma directa o indirecta por la pandemia son mayores. Por lo tanto, el excedente de defunciones es mayor a la comparación de la situación actual con la de años anteriores.

**Referencias**

1. Gobierno del Perú. Decreto Supremo N° 008-2020-SA Decreto Supremo que declara en emergencia sanitaria a nivel nacional por el plazo de noventa (90) días calendario y dicta medidas prevención y control del Covid-19. Diario oficial El Peruano [Internet]. [citado el 7 de junio de 2020]; Disponible en: https://busquedas.elperuano.pe/normaslegales/decreto-supremo-que-declara-en-emergencia-sanitaria-a-nivel-decreto-supremo-n-008-2020-sa-1863981-2/

2. Ministerio de Educación. Resolución Viceministerial N° 084-2020-MINEDU Disponen medidas excepcionales con relación al servicio educativo que se realiza de forma presencial, correspondiente al año lectivo 2020 brindado por los Centros de Educación Técnico-Productiva e Institutos y Escuelas de Educación Superior públicos y privados. Diario oficial El Peruano [Internet]. [citado el 7 de junio de 2020]; Disponible en: https://busquedas.elperuano.pe/normaslegales/disponen-medidas-excepcionales-con-relacion-al-servicio-educ-resolucion-vice-ministerial-n-084-2020-minedu-1865287-1/

3. Gobierno del Perú. Decreto Supremo que declara Estado de Emergencia Nacional por las graves circunstancias que afectan la vida de la Nación a consecuencia del brote del COVID-19-DECRETO SUPREMO-N° 044-2020-PCM. Diario oficial El Peruano [Internet]. [citado el 7 de junio de 2020]; Disponible en: http://busquedas.elperuano.pe/normaslegales/decreto-supremo-que-declara-estado-de-emergencia-nacional-po-decreto-supremo-n-044-2020-pcm-1864948-2/

4. Gobierno del Perú. Decreto Supremo que establece las medidas que debe observar la ciudadanía hacia una nueva convivencia social y prorroga el Estado de Emergencia Nacional por las graves circunstancias que afectan la vida de la Nación a consecuencia del COVID-19-DECRETO SUPREMO-N° 094-2020-PCM. Diario oficial El Peruano [Internet]. [citado el 7 de junio de 2020]; Disponible en: http://busquedas.elperuano.pe/normaslegales/decreto-supremo-que-establece-las-medidas-que-debe-observar-decreto-supremo-n-094-2020-pcm-1866708-1/

5. Google. COVID-19 Community Mobility Report [Internet]. COVID-19 Community Mobility Report. [citado el 7 de junio de 2020]. Disponible en: https://www.google.com/covid19/mobility?hl=es

6. Gobierno del Perú. Coronavirus (COVID-19) en el Perú [Internet]. [citado el 7 de junio de 2020]. Disponible en: https://www.gob.pe/coronavirus

7. Ministerio de Salud. Información de Fallecidos del Sistema Informático Nacional de Defunciones - SINADEF [Internet]. [citado el 7 de junio de 2020]. Disponible en: https://www.datosabiertos.gob.pe/dataset/informaci%C3%B3n-de-fallecidos-del-sistema-inform%C3%A1tico-nacional-de-defunciones-sinadef-ministerio

8. Nogueira PJ, Nobre M de A, Nicola PJ, Furtado C, Carneiro AV. Excess Mortality Estimation During the COVID-19 Pandemic: Preliminary Data from Portugal. Acta Médica Port. el 1 de junio de 2020;33(6):376–83.

9. Weinberger DM, Cohen T, Crawford FW, Mostashari F, Olson D, Pitzer VE, et al. Estimating the early death toll of COVID-19 in the United States. bioRxiv [Internet].

el 18 de abril de 2020 [citado el 25 de junio de 2020]; Disponible en: https://www.ncbi.nlm.nih.gov/pmc/articles/PMC7217085/

10. Ministerio de Salud. SINADEF Sistema Informático Nacional de Defunciones [Internet]. [citado el 7 de junio de 2020]. Disponible en: https://www.minsa.gob.pe/defunciones/

11. Vargas-Herrera J, Pardo Ruiz K, Garro Nuñez G, Miki Ohno J, Pérez-Lu JE, Valdez Huarcaya W, et al. Resultados preliminares del fortalecimiento del sistema informático nacional de defunciones. Rev Peru Med Exp Salud Publica. julio de 2018;35(3):505–14.

12. Hart JD, Sorchik R, Bo KS, Chowdhury HR, Gamage S, Joshi R, et al. Improving medical certification of cause of death: effective strategies and approaches based on experiences from the Data for Health Initiative. BMC Med. el 9 de marzo de 2020;18(1):74.

13. Instituto Nacional de Estadística e Informática (INEI). SIRTOD Sistema de Información Regional para la Toma de Decisiones [Internet]. [citado el 7 de junio de 2020]. Disponible en: http://systems.inei.gob.pe:8080/SIRTOD/app/consulta

14. Millimet DL, Parmeter CF. Accounting for Skewed or One-Sided Measurement Error in the Dependent Variable. IZA Discussion Paper No. 12576 [Internet]. [citado el 7 de junio de 2020]. Disponible SSRN: https://ssrn.com/abstract=3449571

15. Instituto Nacional de Estadística e Informática (INEI). Censos Nacionales 2017: XII de Población, VII de Vivienda y III de Comunidades Indígenas [Internet]. [citado el 7 de junio de 2020]. Disponible en: http://censo2017.inei.gob.pe/metodologia/

16. Instituto Nacional de Estadística e Informática (INEI). Perú: Perfil Sociodemográfico. Informe Nacional [Internet]. [citado el 7 de junio de 2020]. Disponible en: https://www.inei.gob.pe/media/MenuRecursivo/publicaciones_digitales/Est/Lib1539/

17. Choi I. Unit root tests for panel data. J Int Money Finance. 2001;20(2):249–272.

18. Wooldridge JM. Econometric analysis of cross section and panel data. MIT press; 2010.

19. Hausman JA. Specification Tests in Econometrics. Econometrica. 1978;46(6):1251–71.

20. Modi C, Boehm V, Ferraro S, Stein G, Seljak U. How deadly is COVID-19? A rigorous analysis of excess mortality and age-dependent fatality rates in Italy. medRxiv. el 14 de mayo de 2020;2020.04.15.20067074.

21. Bonardi J-P, Gallea Q, Kalanoski D, Lalive R. Fast and local: lockdown policies affect the spread and severity of covid-19. Covid Econ 325. 2020;251.

22. Roux J, Massonnaud C, Crépey P. COVID-19: One-month impact of the French lockdown on the epidemic burden. medRxiv. el 27 de abril de 2020;2020.04.22.20075705.

23. Lau H, Khosrawipour V, Kocbach P, Mikolajczyk A, Schubert J, Bania J, et al. The positive impact of lockdown in Wuhan on containing the COVID-19 outbreak in China. J Travel Med [Internet]. el 18 de mayo de 2020 [citado el 26 de junio de 2020];27(3). Disponible en: https://academic.oup.com/jtm/article/27/3/taaa037/5808003

24. De Figueiredo AM, Codina AD, Cristina D, Figueiredo MM, Saez M, León AC. Impact of lockdown on COVID-19 incidence and mortality in China: an interrupted time series study. Bull World Health Organ. E-pub: 6 de abril de 2020;19.

25. Husain Z, Das AK, Ghosh S. Did the National lockdown lock COVID-19 down in India, and reduce pressure on health infrastructure? medRxiv. el 29 de mayo de 2020;2020.05.27.20115329.

26. Meunier TAJ. Full lockdown policies in Western Europe countries have no evident impacts on the COVID-19 epidemic. medRxiv. el 1 de mayo de 2020;2020.04.24.20078717.

27. He G, Pan Y, Tanaka T. COVID-19, City Lockdowns, and Air Pollution: Evidence from China. medRxiv. el 21 de abril de 2020;2020.03.29.20046649.

28. Bhuiyan AKMI, Sakib N, Pakpour AH, Griffiths MD, Mamun MA. COVID-19-Related Suicides in Bangladesh Due to Lockdown and Economic Factors: Case Study Evidence from Media Reports. Int J Ment Health Addict [Internet]. el 15 de mayo de 2020 [citado el 29 de junio de 2020]; Disponible en: http://link.springer.com/10.1007/s11469-020-00307-y

29. Sood S. Psychological effects of the Coronavirus disease-2019 pandemic. Res Humanit Med Educ. el 1 de abril de 2020;7:23–6.

30. Gobierno del Perú. Decreto Supremo N° 009-2020-SA Decreto Supremo que declara en Emergencia Sanitaria, por el plazo de noventa (90) días calendario, distritos priorizados de la provincia de La Convención, del departamento de Cusco-. Diario oficial El Peruano [Internet]. [citado el 1 de julio de 2020]; Disponible en: https://busquedas.elperuano.pe/normaslegales/decreto-supremo-que-declara-en-emergencia-sanitaria-por-el-decreto-supremo-n-009-2020-sa-1864050-3/